\documentclass[preprint]{aastex}

\shortauthors{Watson et al.}
\shorttitle{Magnetic Fields and the Circular Polarization}

\begin{document}

\title{Irregular Magnetic Fields in Interstellar Clouds and
Variations in the Observed Circular Polarization of Spectral Lines}

\author{W. D. Watson\altaffilmark{1}, D. S. Wiebe\altaffilmark{1,2,3}, and R. M. Crutcher\altaffilmark{2}}
\affil{University of Illinois at Urbana-Champaign, Urbana, IL 61801}
\email{w-watson@uiuc.edu}
\altaffiltext{1}{Department of Physics}
\altaffiltext{2}{Department of Astronomy}
\altaffiltext{3}{Permanent address: Institute of Astronomy of the RAS,
48, Pyatnitskaya str., 109017 Moscow, Russia}

\begin{abstract}
The strengths of magnetic fields in interstellar gas clouds are obtained through
observations of the circular polarization of spectral line radiation. Irregularities in this
magnetic field may be present due to turbulence, waves or perhaps other causes, and may
play an essential role in the structure and evolution of the gas clouds. To infer
information about these irregularities from the observational data, we develop statistical
relationships between the rms values of the irregular component of the magnetic field and spatial
variations in the circular polarization of the spectral line radiation. The irregularities are
characterized in analogy with descriptions of turbulence---by a sum of Fourier waves
having a power spectrum with a slope similar to that of Kolmogorov turbulence. For
comparison, we also perform computations in which turbulent magnetic and velocity
fields from representative MHD simulations by others are utilized.
Although the effects of the variations about the mean value of the magnetic field along
the path of a ray tend to cancel, a significant residual effect in the polarization of the
emergent radiation remains for typical values of the relevant parameters. A map of
observed spectra of the 21 cm line toward Orion A is analyzed and the results are
compared with our calculations in order to infer the strength of the irregular component
of the magnetic field. The rms of the irregular component is found to be comparable in
magnitude to the mean magnetic field within the cloud. Hence, the turbulent and Alfven velocities should also be comparable.
\end{abstract}

\keywords{ISM: clouds---ISM: magnetic fields---polarization---stars: formation---turbulence}

\section{Introduction}

The breadths of spectral line radiation from interstellar gas clouds at radio
wavelengths are recognized to be much greater than would be caused by Doppler shifts
resulting from thermal motions alone at the temperatures of the atoms and molecules in
the gas. These breadths are believed to be an indication of disturbances (most likely a
form of turbulence or waves) which are fundamental to the structure and evolution of
interstellar gas clouds and star formation (e.g., V\'azquez-Semadeni et al. 2000, and reviews in Lada \& Kylafis 1999 and
in Franco \& Carraminana 1999). Although the conclusions of earlier investigations that
relate detailed information about the spectral lines specifically to the properties of
turbulence have been questioned (see Dubinski, Narayan, \& Phillips 1995), recent efforts
along these lines are encouraging (Miesch \& Scalo 1995; Lis et al. 1998). Of particular
importance is the role of waves or turbulence in causing a magnetic pressure that acts
parallel to the direction of the mean magnetic field lines in the clouds. This has long been
viewed as a promising way to understand why the rate of collapse of interstellar clouds
along the field lines is not more rapid than is indicated by the observational data.
Magnetic disturbances have been thought to decay more slowly than
ordinary disturbances, and hence more likely to persist and provide the component of the
pressure that seems to be necessary (Arons \& Max 1975). However, the premise that the magnetic disturbances
are longer lived has been questioned in recent investigations (MacLow et al. 1998;
Stone, Ostriker, \& Gammie 1998; Padoan \& Nordlund 1999). In any case, irregularities
in the magnetic fields are expected to be associated with the bulk motions which are
reflected by the non-thermal breadths of spectral lines. At some level, these should be
evident in the observational data.

Information about the direction of the magnetic field in the interstellar gas has long
been inferred from the linear polarization of starlight that results from selective absorption by dust
grains. However, the polarizing power of the grains that absorb starlight apparently is
greatly reduced in gas clouds of even modest densities (e.g., Goodman 1996). Though
such polarization data are extensive, they do not thus seem to be sensitive indicators for
the irregularities of the magnetic field in dense clouds. In contrast, there is no doubt that
the magnetic field in interstellar clouds is evident in the circular polarization of spectral
lines at radio frequencies through the Zeeman effect. Although such data are not so
extensive as is desirable for extracting information on irregularities in the field, advances
in the observational capabilities are continuing. It is thus useful to examine quantitatively
the sensitivity of the circular polarization of spectral lines to plausible irregularities in the
magnetic field in interstellar clouds. In particular, the polarization associated with a ray
of radiation is a sum of contributions by the gas all along the path of the ray. The effects
of deviations of the field from its mean value thus tend to cancel for the radiation that
emerges from the cloud and to reduce its sensitivity as an indicator of irregularities. The
goal of our investigation is to determine what information remains in the circular
polarization of radiation that emerges from an irregular medium. Though not the
emphasis of our investigation, an important challenge is to identify observational
quantities for which the predictions distinguish between actual MHD phenomena and a
simple generic description of the irregularities in the fields. Finally, we note that the
emission of linearly polarized radiation has been observed from grains in dense clouds.
This linearly polarized radiation holds promise to become a valuable tool (e.g.,
Hildebrand 1996; Rao et al. 1998) as more sensitive and extensive maps are obtained.

In Section 2, we describe how the irregularities are characterized and how the
calculations of the circular polarization are performed. We focus on the quantity that
commonly is obtained from the observational data---the single value $B_{\rm f}$
of the magnetic field that best fits the circular polarization (Stokes $V$ intensity) when the
Zeeman splitting is much smaller than the breadth of the spectral line. By comparing the
statistical properties of the $B_{\rm f}$
which are inferred from the observations with the properties of the $B_{\rm f}$
which we calculate, we propose to gain information about the actual strength of the
irregular component of the field within interstellar clouds. As an example, data for the 21
cm spectral line toward Orion A are analyzed in this manner in Section 3.

\section{Calculations}

The calculations here are restricted to the limit that ordinarily is relevant for
interstellar gas clouds---that in which the Zeeman splitting is much smaller than the
breadth of the spectral line. The optical depth $\tau (v)$
for the intensity of the spectral line as a function of Doppler velocity $v$
can then be expressed as,
\begin{equation}
\tau (v) = \int f(s)\exp(-(v-v_{\rm t})^2/v_{\rm th}^2 )\,ds,
\label{e21}
\end{equation}
where the integral is along a straight-line path of length $s$ through the medium in which
the turbulent velocity $v_{\rm t}$ varies along the path. The integral can be understood
as a sum of
the contributions from components of the gas along the line of sight where the bulk
velocity of the component at location $s$ is $v_{\rm t}(s)$. The exponential arises from the
Maxwellian form for the distribution of atomic/molecular velocities at any location and
depends upon the thermal velocity $v_{\rm th}=\sqrt{2kT/m}$.
All of the necessary information about the excitation, the abundance of the relevant
species, and the atomic/molecular data is incorporated into the opacity function $f(s)$. As
measured by the Stokes $V$ intensity, the net circular polarization is then related to the
difference $\tau_V(v)\ll 1$ between the optical depths associated with the two senses
of circular polarization,
\begin{equation}
\tau_V(v)=-2pv_{\rm th}^{-2}\int f(s)B(s)(v-v_{\rm t})
\exp(-(v-v_{\rm t})^2/v_{\rm th}^2)\,ds,
\end{equation}
where $p$ is a constant that depends upon the particular transition and $B(s)$ is the
component of the magnetic field that is parallel to the path (and hence to the line of sight
of the observer). Note that when $B$ is constant along the path of the radiation,
\begin{equation}
\tau_V=pB\partial\tau(v)/\partial v.
\label{e23}
\end{equation}
If the spectral line in the gas is being observed in absorption in front of a strong
continuum source with intensity $I_0$ as occurs toward Orion A in the analysis in Section 3,
the observed intensity $I$ is given by
\begin{equation}
I=I_0\exp(-\tau)
\end{equation}
and Stokes-$V$ is given by
\begin{equation}
V=\tau_VI,
\end{equation}
so that if $B$ were constant
\begin{equation}
V = pB\partial I/\partial v
\label{e26}
\end{equation}
and $B$ could readily be determined from the observation of $I$ and $V$. Even though the
magnetic field may not be constant along the path of the ray of radiation, a ``best fit'' (in
the least squares sense) value $B_{\rm f}$ often is obtained from equation (\ref{e23}),
\begin{equation}
B_{\rm f}=\int(\tau_V\partial\tau/\partial v)dv/
\int p(\partial\tau/\partial v)^2\,dv
\label{e27}
\end{equation}
or by the analogous fit to equation (\ref{e26}).

If there are irregularities in the magnetic field, the best fit values for $B_{\rm f}$
will not be the same for spectral lines that are emitted from different locations across the
surface of the cloud. Even at a single location, irregularities will usually prevent equation
(\ref{e23}) or (\ref{e26}) from being satisfied by a single value for the magnetic field at all Doppler
velocities within the observed spectral line. These variations in the inferred $B_{\rm f}$
are related, though only indirectly through the emergent Stokes-$V$, to the variations in the
magnetic field $B(s)$ along the path of the radiation. The purpose of the investigation here
is to provide a statistical relationship between the variations in the values of $B_{\rm f}$
that are inferred from the observational data and the amplitudes of the actual
irregularities in the magnetic fields that are the cause of such variations. While more
sophisticated and involved analytical tools might be imagined, we will limit our attention
here to two basic quantities. (i) The standard deviation of the values of $B_{\rm f}$
that are inferred from the radiation emerging at the various different locations which
comprise the surface of the cloud
\begin{equation}
(B_{\rm f})_{\rm sd}=\left\{\left\langle B_{\rm f}^2\right\rangle-
\left\langle B_{\rm f}\right\rangle^2\right\}^{1/2}.
\end{equation}
(ii) The average across the surface of the cloud of the residuals
$\langle(B_{\rm res}^2)^{1/2}\rangle$ that remain when best fits are obtained for
$B_{\rm f}$ at the various locations where at a single location
\begin{equation}
B_{\rm res}^2=\int(\tau_V-pB_{\rm f}\partial\tau/\partial v)^2 dv/
\int(p\partial\tau/\partial v)^2\,dv.
\end{equation}
Irregularities in the velocity and magnetic fields in the gas are characterized here
in a conventional manner by the exponent in the power law distribution for the
amplitudes of the Fourier components, by the root mean square (rms) of the velocity and
of the magnetic fields, and by the smallest wave number (corresponding to the largest
scale length) to which the power law distribution extends. For turbulence, this scale
length represents the largest scale length at which disturbances are injected into the
gas. The scale length implies a ``correlation length'', which we also use in describing
the calculations. Representative configurations (or statistical ``realizations'') for the
fields
are then created by statistical sampling with Gaussian distributions for the Fourier
amplitudes. Such methods are standard and are described, for example, in our own
previous investigations (Wallin, Watson, \& Wyld 1998, 1999) as well as elsewhere (e.g.,
Dubinski, Narayan, \& Phillips 1995) where turbulent velocity fields are created. The
turbulent magnetic fields are created in the same way (e.g., Wiebe \& Watson
1998). For comparison purposes, we present a few examples of the realizations created
by statistical sampling where the slope of the power spectrum is exactly that given by the
Kolmogorov exponent. However, we focus our main attention on calculations in which
the spectrum is somewhat steeper. This steeper spectrum leads to velocities and magnetic
fields in coordinate space for which the slope of the correlation function is in better
agreement with that calculated utilizing the fields from the available MHD simulations
(see below). We have computed the ``structure function'' for the magnetic and velocity
fields, as well as for the mass density, from these MHD computations. We find that the
variation of the structure function (as defined by, e.g., Frisch 1995) corresponds best to
(separation)$^{4/3}$, whereas the Kolmogorov variation is (separation)$^{2/3}$. The variation of the structure function is related to the variation of the power spectrum (e.g., Frisch 1995). We thus
focus on a distribution for the power spectrum in Fourier space in our statistical
sampling $(k^{-7/3})$ that is somewhat steeper than that of Kolmogorov $(k^{-5/3})$. We have
verified that the structure functions which are computed using the fields obtained from
the steeper power spectrum do agree well with those that are computed directly from the
fields of the MHD simulations. We note that the $(k^{-7/3})$ dependence that we infer may not be a basic property of MHD turbulence, but may be a result of the numerical resolution in the MHD computations and the way in which the disturbances are injected. The
representative configurations for the velocities and magnetic fields that are created
according to the foregoing statistical procedure thus provide a ``generic'' description for
the irregular fields in terms of a minimal set of basic parameters.

     In detail, the foregoing generic description certainly is
incomplete---most notably, it incorporates no correlation between the velocities and
magnetic fields
and it contains no information on the variations in the density of matter. To obtain a
quantitative measure of the influence of these deficiencies, we also perform computations
in which these quantities are provided by the results of numerical simulations by others
for time-independent, compressible MHD turbulence. We refer the reader to the papers
by these authors for a detailed description of the calculational methods (see Stone,
Ostriker, \& Gammie 1998). Certain basic aspects of these calculations should, however, be summarized. The turbulence is driven by
continuously injecting disturbances at a range of scale lengths centered on one-sixth the
length of an edge of the computational cube. This determines the minimum wavenumber
in the spectrum of the turbulence, and hence the correlation length for the velocities. In
the way that we measure the correlation length (see below), the resulting correlation
length corresponds to about one-twelfth the length of the edge of the computational cube.
Three cases are considered according to the relative strength of the mean magnetic field
$B_{\rm avg}$ and the thermal gas pressure $P_{\rm th}$ as described by the usual ``plasma
parameter'' $\beta=(8\pi P_{\rm th}/B_{\rm avg}^2 )=$ 0.02, 0.2 and 2. For convenience, we label these as S, M, and W for strong, medium and weak, respectively. For
all three cases, the rms turbulent velocity in each of the three orthogonal directions is close to the
value of $2.9c_{\rm s}$ in the weak (W) field case where the velocities are essentially
isotropic.
With increasing strength of the mean magnetic field, the gas becomes somewhat
anisotropic so that the rms turbulent velocities parallel and perpendicular to the mean
field are $3.4c_{\rm s}$ and $2.8c_{\rm s}$, respectively, in the strong (S) case. Although we are limited to
essentially a single ratio of the turbulent velocity to the sound velocity in the available
MHD simulations, this ratio of approximately three is representative for relevant
interstellar clouds. The chief difference between the three MHD
cases is in the relative strengths of the mean magnetic fields which are in the ratio
1(W):3.2(M):10(S)---assuming that the same $P_{\rm th}$ is adopted in all three cases. The ratio of the standard deviation of the magnetic field in each of the three orthogonal directions to the mean
magnetic field in these simulations is approximately 5/3 (W case), 2/3 (M case), and 0.15 [parallel]/0.25
[perpendicular] (S case). These ratios can be viewed as depending mainly on the Alfvenic Mach number---the ratio of the three dimensional rms turbulent velocity to the Alfven velocity. The Alfvenic Mach number is 5(W), 1.6(M) and 0.5(S) in the simulations (the sonic Mach number is approximately five in all three simulations). The turbulent magnetic fields are then similar in the three cases,
as are the turbulent velocities. To
relate the dimensionless results of the MHD computations to physical units, it is useful
to recognize
that there is freedom to choose the values for two additional scaling factors after $\beta$ has
been specified. These can be the sound velocity $c_{\rm s}$ and the average matter density.
Representative values for molecular clouds (e.g., Crutcher 1999) are
$\beta\simeq0.01$ to 0.1 and include
$3c_{\rm s}$ within the range of rms turbulent velocities that are observed.
 Their gas
temperatures typically are 10 to 30K, though the Orion A cloud that is analyzed in
Section 3 is somewhat warmer ($\simeq100$~K).

Two idealizations will be considered for the opacity function $f(s)$---(i) $f(s)$ will be
taken as a constant, and (ii) $f(s)$ will be taken as proportional to the matter density. The
latter case is considered only in conjunction with the fields from the MHD simulations
where we have information about the variations in the matter density.

\begin{figure}
\plotone{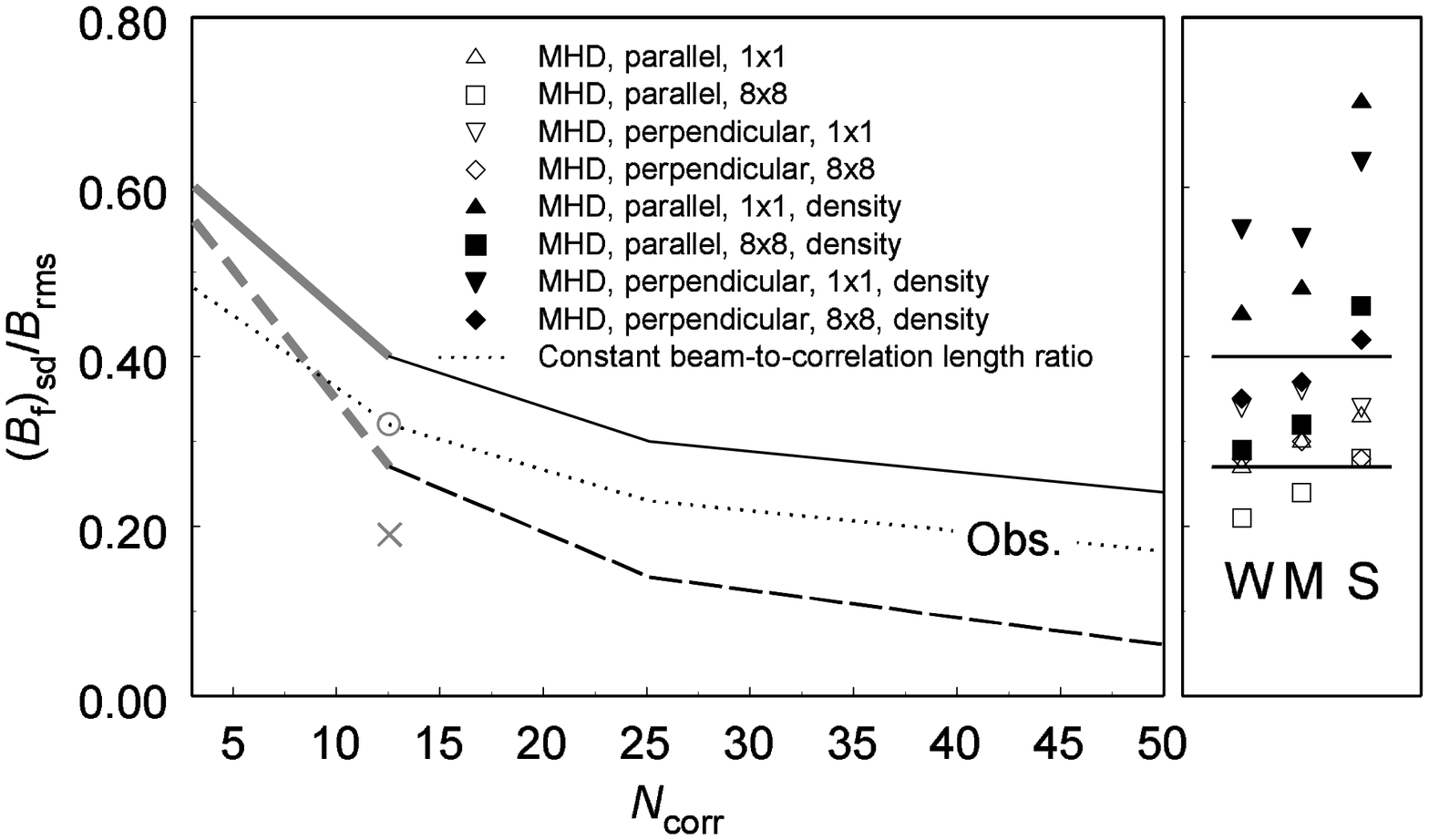}
\caption{%
(left-hand panel) The ratio $(B_{\rm f} )_{\rm sd} /B_{\rm rms}$
versus the number $N_{\rm corr}$
 of correlation lengths across the edge of the computational cube---based on fields created
by statistical sampling. The uppermost (solid) line is computed when each of the (128)$^2$ rays is treated independently. The lowest (dashed) line is computed from the (16)$^2$ values of
$B_{\rm f}$ that are obtained when the radiation from areas that are 8 grid points on a
side is combined before $B_{\rm f}$
is determined. The open circle and the cross represent computations in which the
Kolmogorov exponent is used for the power spectrum. As discussed in the text, a
somewhat steeper power spectrum is utilized in the other computations for this panel.
The segments that are indicated by broader lines are especially dependent upon the
particular statistical realization of the fields. In this region, the curves represent the average from several
realizations. The line labeled ``Obs'' is intended to incorporate the finite angular size of the beam for the observational data which are analyzed in Section 3.
(right-hand panel) The ratio $(B_{\rm f} )_{\rm sd} /B_{\rm rms} $
computed with the fields obtained from MHD simulations for the cases in which the average
magnetic field is weak (W), medium (M), and strong (S). This ratio is presented for
computations in which the radiation propagates parallel or perpendicular to the direction
of the non-zero average magnetic field, both for individual rays (1x1) and for areas of 8
grid points on a side (8x8) as in the left-hand panel. The opacity function $f(s)$ is
considered to be either independent of mass density (open symbols) or to be proportional
to the mass density (filled symbols). To facilitate comparisons, the two horizontal lines in
this panel indicate the values of $(B_{\rm f} )_{\rm sd} /B_{\rm rms}$
taken from the left-hand panel for individual rays and for areas of 8x8 grid points.
}
\label{fig1}
\end{figure}

\placefigure{fig1}

In Figure 1, we show the results for the standard deviations $(B_{\rm f})_{\rm sd}$
that are obtained from the computed Stokes-$V$ as a function of the various parameters.
For the statistically created fields, $(B_{\rm f})_{\rm sd}$
is independent of the average magnetic field in the gas. The results can thus be
presented usefully in terms of the ratio $(B_{\rm f})_{\rm sd}/B_{\rm rms}$
where $B_{\rm rms}$ is the rms of the irregular (or turbulent) component of the magnetic
field in the gas. That is, $B(s)=B_{\rm avg}+ B_{\rm irregular}$
for which $B_{\rm rms}\equiv \left\langle {B_{\rm irregular}^2 }\right\rangle^{1/2}$.
Calculations are presented when $B_{\rm f}$
is determined separately for each of the (128)$^2$ rays that emerge from the grid points on the surface
and perpendicular to the surface of our computational cube consisting of (128)$^3$ grid
points. Calculations also are presented when the intensity and Stokes-$V$ for rays from
separate areas of 8x8 surface grid points are first summed before the $B_{\rm f}$
are obtained from equation (\ref{e27}). The latter is intended to provide an indication of the
influence of the finite angular resolution of a telescope. The calculations are performed as
a function of the minimum wavenumber (our parameter $k_{\min}$
as defined in Wallin et al. 1998) at which the irregularities are injected into the medium.
The correlation length is then computed from the structure function of the resulting fields
in coordinate (that is, ordinary physical) space. We define the correlation length here as
the separation at which the structure function ``becomes flat''. Due (presumably) to
computational imprecision, the structure function does not become completely flat but
has a wavelike variation with an amplitude of a few percent. We thus take the correlation
length to be the separation at which the structure function becomes flat after this wave
has been subtracted. In Figure 1, the quantity
$(B_{\rm f})_{\rm sd}/B_{\rm rms}$ is then presented as a function of the number
$N_{\rm corr}$
of such correlation lengths along an edge of the computational cube. Other than those at a single correlation length corresponding to
$N_{\rm corr}=12$, all of the computations are performed with fields obtained from the power spectrum $(k^{-7/3})$ that is somewhat steeper than Kolmogorov. Within the range of turbulent velocities in our computations
(see Figure 2) which covers
the range for interstellar clouds, any dependence of $(B_{\rm f})_{\rm sd} /B_{\rm rms}$
upon the rms value of the turbulent velocities is negligible. For $N_{\rm corr}\lesssim12$,
the longest wavelengths in the turbulence are an appreciable fraction of the size of the
cube and the purely statistical variations in $(B_{\rm f})_{\rm sd} /B_{\rm rms}$
from one realization of the cube to another (with otherwise identical parameters) are
appreciable. This is indicated in Figure 1 by the broader lines. These lines are obtained by
averaging $(B_{\rm f})_{\rm sd}/B_{\rm rms}$ for several statistical realizations.
For $N_{\rm corr}\gtrsim12$, the variations in $(B_{\rm f})_{\rm sd} /B_{\rm rms}$
from one statistical realization to another are negligible.

In the right-hand panel in Figure 1, we present $(B_{\rm f})_{\rm sd}/B_{\rm rms}$
obtained by utilizing fields from the MHD computations of Stone et al. (1998). These computations are available only for a
single correlation length which corresponds to $N_{\rm corr}\simeq12$ because the turbulence in
these computations always is injected with the same distribution of scale lengths. In the
strong field case, the correlation length is somewhat different in the directions
perpendicular to the average magnetic field. For ease of comparison, the
$(B_{\rm f})_{\rm sd}/B_{\rm rms}$
based on the statistically created fields for $N_{\rm corr}=12$
in the left-hand panel are indicated by the two horizontal lines in the right-hand panel. In
addition to the turbulent magnetic and velocity fields, there is a non-zero average
magnetic field in the MHD simulations. This influences the medium, and potentially the
results of our computations (a non-zero average magnetic field has no effect on the
results of computations in the left-hand panel with the statistically created fields). As
noted previously, observations suggest that molecular clouds fall between
the medium and strong field cases in terms of the plasma parameter $\beta$. There is no reliable way
to incorporate variations in the mass density into the computations that utilize the
statistical fields. Direct comparisons can thus only be made with the  computations
for the MHD fields where variations in the mass density are ignored (open symbols) in
computing the optical depths. The computations in which $f(s)$ in equation (\ref{e21}) is
proportional to the mass density are indicated by filled symbols. For the computations
with the finite (8x8) surface areas and when the optical depths do not depend upon mass
density, $(B_{\rm f})_{\rm sd}/B_{\rm rms}$
ranges from approximately 0.25 to 0.30 in the case of medium (M) field strength
according to whether the line of sight is parallel or perpendicular to the direction of the
average magnetic field. For the analogous correlation length $(N_{\rm corr}= 12)$,
the computation with the statistically created fields is midway between these values.
When the optical depths are assumed to be proportional to mass density,
$(B_{\rm f} )_{\rm sd}/B_{\rm rms} $ can be seen to range from 0.33 to 0.39 when the same
fields from the MHD computations are utilized.

\begin{figure}
\plotone{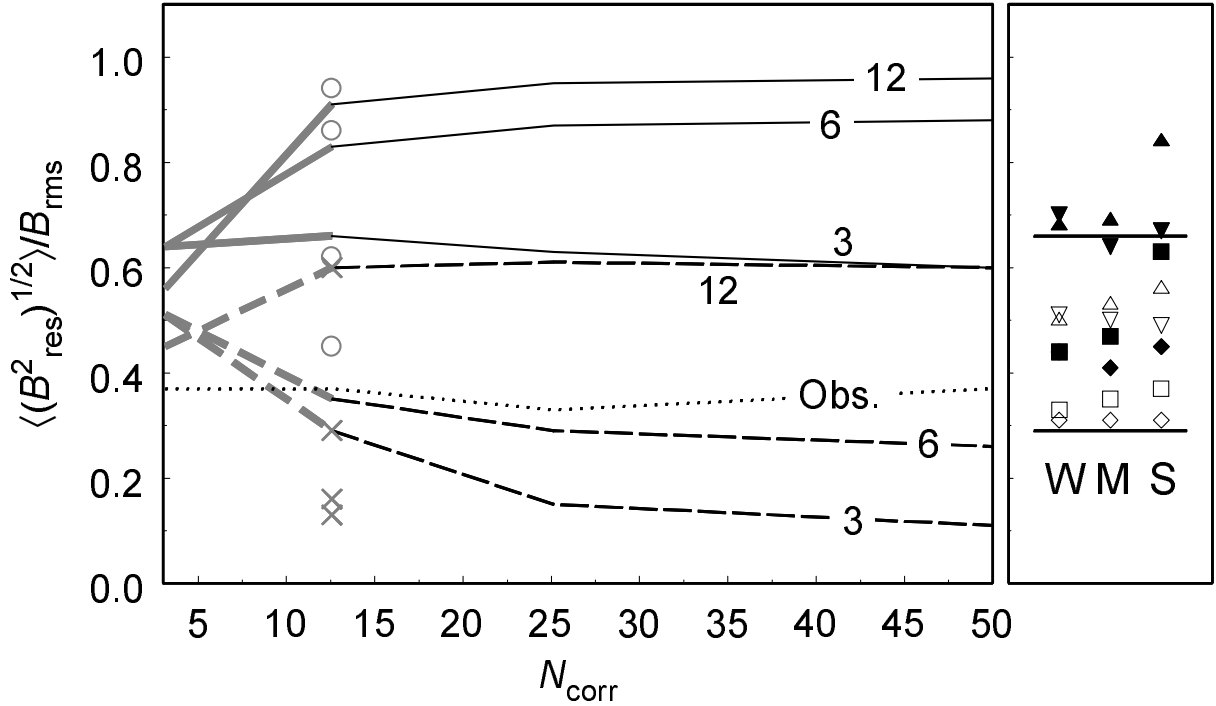}
\caption{%
(left-hand panel) The ratio $\langle (B_{\rm res}^2 )^{1/2} \rangle /B_{\rm rms} $
 versus the number $N_{\rm corr}$
 of correlation lengths across the edge of the computational cube---based on fields created
by statistical sampling. The curves are labeled by the value of the rms turbulent velocity
in units of $v_{\rm th}$. The upper three (solid) lines are for the (128)$^2$ rays treated independently. The
lower three (dashed) lines are computed for the (16)$^2$ values of $B_{\rm res}^2 $ that are
 obtained when the radiation within areas that are 8 grid points on a side is combined
before $B_{\rm res}^2 $
 is computed. Other aspects of the Figure have meanings similar to their meanings in
Figure 1. The broader segments of the lines indicate that, as in Figure 1, the results are
sensitive to the particular statistical realization of the fields in this region. The open
circles and the crosses represent computations in which the Kolmogorov value for the
exponent is used for the power spectrum. As in Figure 1, a somewhat steeper power
spectrum is utilized in the other computations for this panel. The lowest circle and cross
are computed with a rms turbulent velocity that is similar to that of the MHD
calculations. The line labeled ``Obs'' is intended to incorporate the finite angular size of the beam for the observational data which are analyzed in Section 3.
(right-hand panel) The ratio $\langle (B_{\rm res}^2 )^{1/2} \rangle /B_{\rm rms} $
computed for fields obtained from the MHD simulations for the cases when the average
magnetic field is weak (W), medium (M), and strong (S). Symbols have the same
meaning as in Figure 1. To facilitate comparisons, the two horizontal lines in this panel
again indicate the value of this ratio when statistically created fields are used, and when the
computations are performed separately for each of the (128)$^2$ rays and when the radiation from areas of 8x8 grid points is first combined. These
are computed with a rms turbulent velocity (2.05 in our units) which is similar to that of
the MHD computations.
}
\label{fig2}
\end{figure}

\placefigure{fig2}

In Figure 2, we show the ratio $\langle (B_{\rm res}^2 )^{1/2} \rangle /B_{\rm rms} $
 calculated with fields that are similar to those used for Figure 1. Unlike the results in
Figure 1, $\langle (B_{\rm res}^2 )^{1/2} \rangle /B_{\rm rms} $
does depend upon the magnitude of the turbulent velocities. Thus, the curves in the
left-hand panel in Figure 2 are labeled according to the ratio of the rms turbulent
velocity to
the rms thermal velocity. In addition to the curves which are computed with the steeper
power spectrum, calculations are again presented for $N_{\rm corr}  = 12$
that utilize the Kolmogorov value for the exponent in the power spectrum (circles and
crosses). As in Figure 1, the right-hand panel shows
$\langle (B_{\rm res}^2)^{1/2} \rangle/B_{\rm rms}$
computed with the MHD fields. The turbulent velocities of the MHD fields are slightly
different from our choices for the curves in the left-hand panel. The results of the
computations with the statistically created fields which are presented for purposes of
comparison in the right-hand panel (horizontal lines) are thus computed with an rms
turbulent velocity that is not exactly the same as that of any of the curves in the left-hand
panel. The agreement between the results of the computations with the statistical and with the MHD fields is generally good. The calculations that are intended to reflect a
finite (8x8) angular resolution coincide for the statistical and MHD fields when the
optical depths do not involve mass density. When the optical depths depend upon mass
density (specifically, when $f$ is proportional to density), $\langle (B_{\rm res}^2 )^{1/2}
\rangle /B_{\rm rms} $
is seen to be increased for the 8x8 resolution from about 0.24 to about 0.30--0.33 for
medium fields. The increase is greater for the strong fields. Because $B_{\rm rms} $
is much less than $B_{\rm avg} $
in strong field case, we suspect that these MHD fields are not so representative of
interstellar clouds where these fields probably are more similar in strength.

\section{An Application to Observational Data}

A statistical analysis such as the foregoing requires measurements of
Stokes-$V$ in the
radiation from a large number of locations on the surface of an
interstellar gas cloud. There is only one existing data set that meets this
requirement---an \ion{H}{1} absorption-line Stokes $I$ and $V$ map across the
surface of a region toward Orion A
(T. H. Troland, R. M. Crutcher, D. A. Roberts, W. M. Goss, \& C. Brogan, in
preparation).
The observations were performed with the VLA in the ``C''
configuration resulting in a spatial resolution of approximately 0.04 pc
(15 arcseconds). The VLA synthesis mapping technique provides uniform
sensitivity to Stokes $I$ and $V$ over the mapped region with no missing
spatial points, which is a desirable characteristic for our analysis.
Although observational data are available for a larger region, we limit the
analysis to the region in which the background source is strong enough that
the detections of the magnetic field are generally at a ``three sigma''
confidence level or better. With the angular resolution of the instrument,
the region that meets our criteria for analysis consists of approximately
100 non-overlapping beams. Hence, we can obtain that number of independent
observational values for $B_{\rm f}$ and $B_{\rm res}^2 $ as a statistical
sample for analysis. Within this area, there are no positions where the
magnetic field was not detected at our sensitivity cutoff. Thus, the data
set has no bias for positions with stronger magnetic fields or against
positions with weaker magnetic fields.
However, this data set does have significant limitations for an
astrophysically meaningful analysis. Toward Orion A the 21 cm spectrum
consists of a pair of partially overlapping absorption lines as indicated
in the spectral data displayed in Figure 3 from a representative location
on the surface. These two \ion{H}{1} velocity components arise in a feature
which has been called the Orion lid. An extensive analysis of \ion{H}{1}
absorption due to the Orion lid has been carried out by van der Werf \&
Goss (1989) based on earlier Stokes $I$ mapping. It consists primarily of
neutral, atomic gas that probably is located close to and on our side of
the \ion{H}{2} region Orion A. The lid extends over at least 1.6 pc in the
plane of the sky. Van der Werf \& Goss argue that the more negative
velocity \ion{H}{1} is primarily photodissociated H$_2$, some of which has been
shocked, and that the more positive velocity component arises in the near
envelope of the molecular cloud that is behind the \ion{H}{2} region. We only
utilize data from the positive velocity side of the intensity minimum that
occurs in Figure 3 at approximately 5 km s$^{-1}$. This is primarily
because the line shape of the other component leads to typically a factor
of two lower sensitivity to $B_{\rm f}$, so fewer positions are available
for statistical analysis. In addition, the more positive velocity component
is likely to be more relevant to the general interstellar medium, and we
wish to avoid confusion in the analysis due to the overlap of these two
components. Specifically, only spectral channels on the positive velocity
side and where the
absorption is less than ninety percent of the maximum absorption are
included. The data give only the optical depth of the \ion{H}{1} line. In order to
infer column densities, one must know the spin temperature of the line.
Toward the main exciting star of the \ion{H}{2} region, $\theta^1$C Orioni, the
\ion{H}{1} column density has been measured with the ultraviolet Lyman-$\alpha$
absorption line; when compared with the 21-cm line optical depth, this
yields a spin temperature of about 100 K. However, there is no information
about possible variations in spin temperature over the mapped area. There
is only the upper limit H$_2$/\ion{H}{1} $<10^{-4}$ available; at least toward
$\theta^1$C Orioni, the Orion lid is not a molecular cloud. Also, there is
no direct information about the volume density of \ion{H}{1}. An assumption of a
spherical cloud and the fact that absorption is seen over at least 1.6 pc
in the plane of the sky leads to $n_{\rm H} \sim 10^3$ cm$^{-3}$. However, it
seems likely that the Orion lid has a more sheet-like geometry, which would
raise the volume density, but by an unknown amount. Because of these
limitations, our analysis of this Orion A data set should be considered to
be primarily an example of the application of our analysis technique rather
than one from which firm astrophysical conclusions should be drawn.

\begin{figure}
\plotone{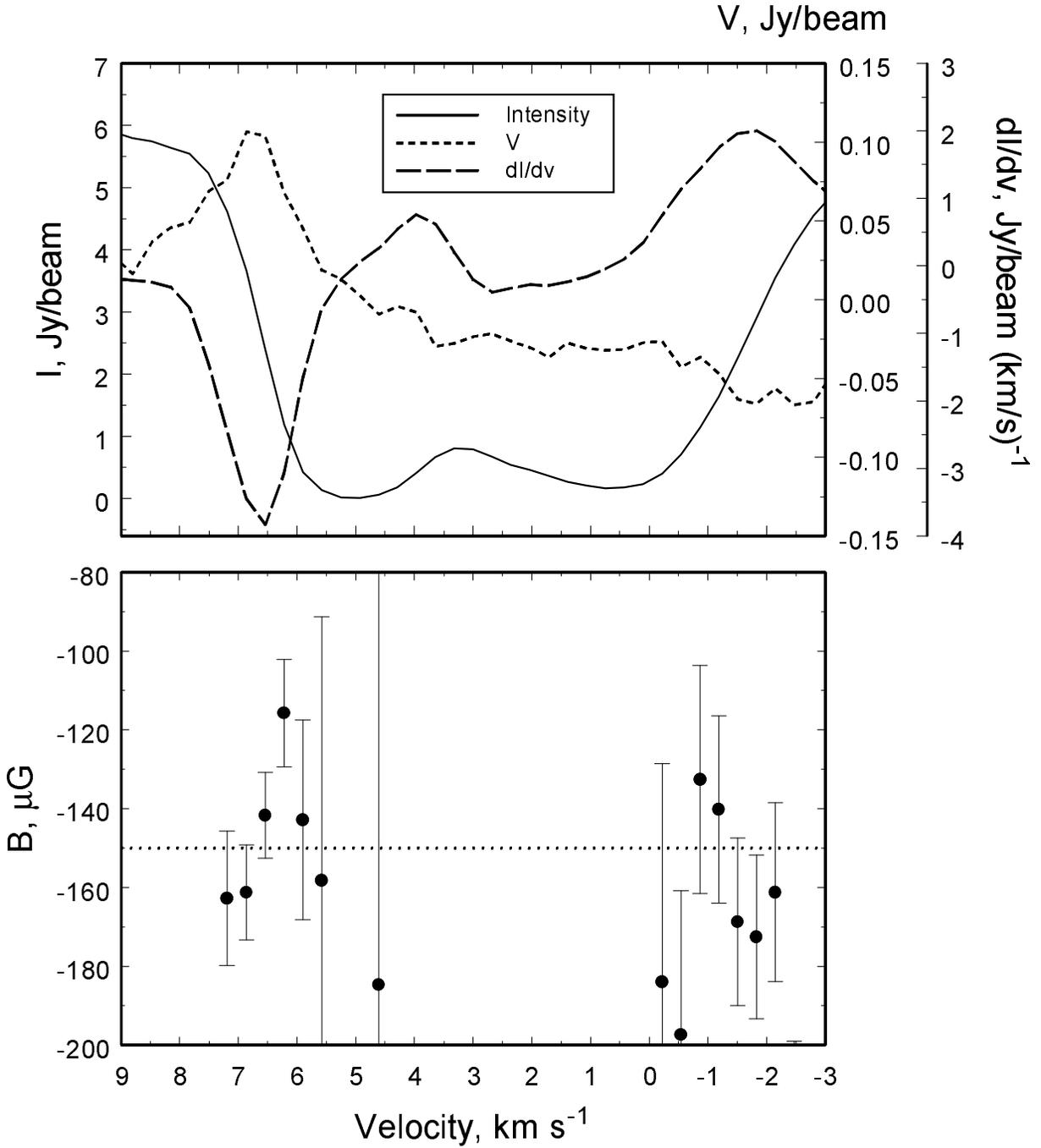}
\caption{%
(upper panel) The observed intensity (Jy per telescope beam), Stokes $V$
intensity (Jy per telescope beam)  and the derivative $dI/dv$ (Jy per km s$^{-1}$ per telescope beam)
of the intensity from a representative location on the surface of Orion A that is included
in our analysis, as function of Doppler velocity.
(lower panel) The values for the magnetic field ($\mu$G) inferred from $B = V/(pdI/dv)$ from the spectrum in the upper panel at
several Doppler velocities. Error bars that correspond to one standard deviation are
shown.
}
\label{fig3}
\end{figure}

\placefigure{fig3}

The formal mean square error associated with the least squares fit to obtain $B_{\rm f}$
in equation (\ref{e27}) from a single, observed spectral line profile is
\begin{equation}
\sigma_{\rm f}^2  = B_{\rm res}^2 /(N - 1),
\end{equation}
where $N$ is the number of intensities within the spectral line that are used to find the value
of $B_{\rm f}$
from that spectral line. For the spectra included in Figure 3, the confidence level thus
corresponds to $B_{\rm f}\ge3\sigma_{\rm f}$.

Differences between Stokes-$V$ and $\partial I/\partial v$
that lead to $\sigma_{\rm f}^2$ can result from noise in the background source and
in the instruments, as well as from
the irregularities in the magnetic field that we are seeking to understand. In the relevant
regime ($\tau_V\ll1$ and $V/I\ll1$), the noise makes a contribution to $\sigma _f^2 $
 that can be expressed as,
\begin{equation}
\sigma_{\rm th}^2\simeq\sum\limits_i {\sigma_{V,{\rm th}}^2 }
(\partial B_{\rm f} /\partial V_i )^2,
\end{equation}
where $\sigma_{V,{\rm th}}^2$
is the mean square variation in the measured $V$ that would occur even if the magnetic
field were constant along the path of a ray. The sum is over the index $i$ that designates the
Stokes parameters $V_i$ which are used in the computation of a particular value of $B_{\rm f}$.
The quantity $\sigma_{V,{\rm th}}^2$ is computed for an adjacent interval of velocities that
is the same size as that used in
computing $B_{\rm f}$, but which is located outside of the spectral line. At these velocities,
measurements of $V$ are not influenced by variations in the magnetic field whereas the variations due to the
noise should be essentially the same as within the spectral line. We thus utilize
\begin{equation}
\sigma_{V,{\rm th}}^2  = \sum\limits_j {V_j^2 } /(N - 1),
\end{equation}
where the index $j$ now designates locations in velocity that are spaced equivalently to
those that are used in computing $B_{\rm f}$, but which are located in the interval
outside of the spectral line. The quantities $\sigma_{\rm f}^2$ and $\sigma _{\rm th}^2$
are computed for each profile and are then averaged over the face of the cloud. We find
for these data toward Orion A
\begin{equation}
\sigma_{\rm th}^2/\sigma_{\rm f}^2  \simeq 0.1.
\end{equation}
Since $\sigma_{\rm th}^2 /\sigma_{\rm f}^2$
is relatively small, for simplicity we will ignore the contribution of the noise in our
further discussions and make the simplifying approximation that $\sigma_{\rm f}$
and $(B_{\rm f})_{\rm sd}$ are due entirely to variations in the magnetic field.

From these observational data toward Orion~A, we find (in agreement with Troland et
al.) for the mean value of $B_{\rm f}$
obtained by averaging over the approximately 100 independent telescope beams that are
included in our analysis
\begin{equation}
\left\langle {B_{\rm f} } \right\rangle ^{\rm obs}  =  - 170\mu{\rm G},
\end{equation}
for the standard deviation of these values
\begin{equation}
(B_{\rm f} )_{\rm sd}^{\rm obs}  = 130\mu {\rm G},
\end{equation}
and for the average of the residuals
\begin{equation}
\langle (B_{\rm res}^2 )^{1/2} \rangle ^{\rm obs}  = 83\mu{\rm G}.
\end{equation}
A histogram of the approximately 100 values of ${B_{\rm f} }^{\rm obs}$ used to find
$\left\langle {B_{\rm f} } \right\rangle ^{\rm obs}$ is given in Figure 4.

\begin{figure}
\plotone{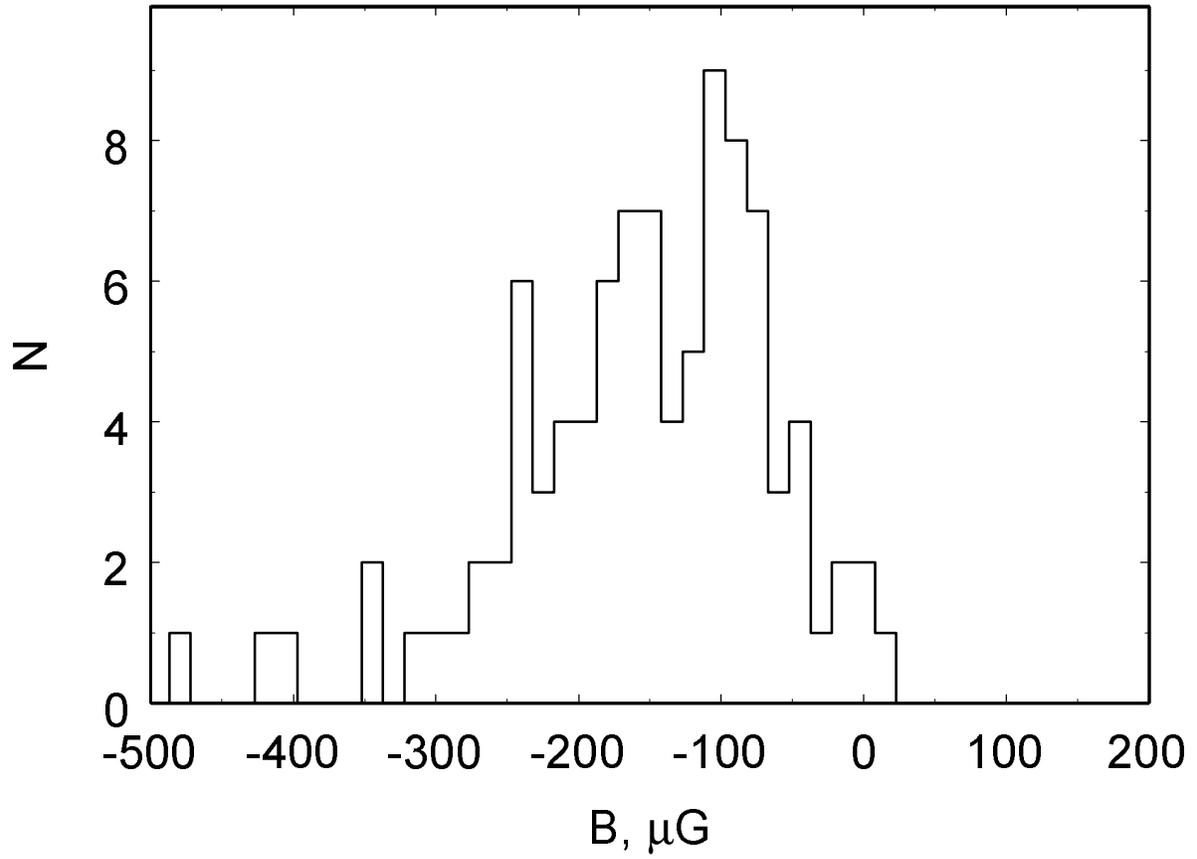}
\caption{%
Histogram of the observationally determined values of ${B_{\rm f} }
^{\rm obs}$ in the region toward Orion A that is being utilized for our statistical analysis.
}
\label{fig4}
\end{figure}

\placefigure{fig4}

To apply our calculations in Figures 1 and 2 for the purpose of inferring the strength
of the irregular component of the actual magnetic field in the Orion~A cloud as
parameterized by $B_{\rm rms}$, the size of the actual gas cloud measured in correlation
lengths  $N_{\rm corr}$
enters. Of course, real interstellar gas clouds are unlikely to be statistically uniform
entities as are our idealized cubes. We can nevertheless estimate a correlation length from
the variation of the measured quantities across the face of the Orion A cloud. That is, the
structure function is computed for the 21 cm line and a correlation length is
inferred in the same way as was done for our theoretical velocities in obtaining Figures 1
and 2. In this way, we infer that there are approximately ten correlation lengths across the
face of the region that is included in our analysis. The size of each telescope beam then
represents about one correlation length. Hence, the relationship between the beam size
and the correlation length is known, but the relevant size of the cloud (which is needed to
obtain $N_{\rm corr}$) is uncertain. Observations indicate only that the face of the entire
cloud is greater than
about 1.6 pc or about forty correlation lengths. The dimension of the cloud along the line
of sight is unknown, but most likely is considerably smaller. Computations are thus
performed for the quantities in Figures 1 and 2 by first summing the intensities within
non-overlapping surface areas that are one correlation length on a side. These are then
averaged over the surface of the cube. Previously, we described first summing over fixed
areas of the surface of the cube to obtain the curves for ``8x8 rays'' in these Figures. Now
the sum is over a number of rays which varies with $N_{\rm corr}$. This seems to be
the most useful way to relate our calculations to the available observational data
for Orion~A. The results are represented by the lines labeled ``Obs'' in Figures 1 and 2.
The ``Obs'' curve in Figure 1 is somewhat less sensitive to the uncertainty about
$N_{\rm corr}$
than are the other curves. In Figure 2, the ``Obs'' curve is quite insensitive to $N_{\rm corr}$
and has a value that is close to 0.4 over the entire range in Figure 2.
The actual $B_{\rm rms}$
that is then implied by the value of 83~$\mu$G for
$\langle (B_{\rm res}^2 )^{1/2} \rangle ^{\rm obs}$ is
\begin{equation}
B_{\rm rms}  \simeq 210\mu{\rm G},
\end{equation}
which indicates that $B_{\rm rms}\simeq \left|{\left\langle {B_{\rm f} }
\right\rangle ^{\rm obs}}\right|$. The observed value
$(B_{\rm f} )_{\rm sd}^{\rm obs}= 130\mu$G, together with the ``Obs'' curve in Figure 1,
seem to imply a similar, though somewhat larger, value for $B_{\rm rms}$. The ordinate for
the  ``Obs'' curve ranges from approximately 0.35 at $N_{\rm corr}=10$
to slightly less than 0.2 at $N_{\rm corr}  = 50$. If $N_{\rm corr}\ge 10$
as seems most appropriate, the inferred random magnetic field is
\begin{equation}
B_{\rm rms}  \gtrsim 370\mu{\rm G}.
\end{equation}
In view of the involved procedures that are used in these two assessments of
$B_{\rm rms}$, the apparent discrepancy is not surprising. If the optical
depths increase with
mass density, the MHD calculations in Figures 1 and 2 indicate that the sense will be to raise the
calculated values in both Figures. The increase will be a larger fraction in Figure 1 than in
Figure 2---an adjustment that is in the correct direction to reduce the discrepancy between
the values for $B_{\rm rms}$
that are inferred in the two approaches. An adjustment of this nature will also decrease the value
for $B_{\rm rms}$
that is inferred from $\langle (B_{\rm res}^2 )^{1/2} \rangle ^{\rm obs}$
to about 130 $\mu$G if one simply scales linearly from the results in the right hand panel in
Figure 2. Our best estimates thus yield the conclusion that $B_{\rm rms} $
is similar in magnitude to the mean value of the magnetic field $\left\langle {B_{\rm f} }
\right\rangle ^{\rm obs}$ for this gas toward Orion A.

We cannot exclude quantitatively the possibility that large scale changes in the
average magnetic field make a significant contribution to the variations that are inferred
for the magnetic field in the Orion A cloud. An indication that such changes are small in
directions along the line of sight is evident in Figure 3. The mean values of the magnetic
fields are essentially the same in the feature on the left-hand side (the feature that we are
analyzing) and in the feature on right-hand side of the profile. Since these two features
probably are separated along the line of sight, we can conclude that the magnetic field
probably does not change significantly in this direction. Deviations between nearby
channels---which presumably are due to changes over much shorter distances---tend to be
greater than the difference between the mean values in the two features. The Troland et
al. (in preparation) map of $B_{\rm f}$
does show evidence of some systematic gradients across the map, though these variations
are not simple and do not appear to be large enough to dominate. Note that $\langle
(B_{\rm res}^2 )^{1/2} \rangle /B_{\rm rms} $
is sensitive to changes in the average magnetic field along the line of sight, but not to
changes across the face of the cloud. In contrast, $(B_{\rm f} )_{\rm sd} /B_{\rm rms} $
is sensitive to changes that occur in directions across the face of the cloud, but not in the
direction along the line of sight.

\section{Discussion}

A number of deficiencies can be imagined in the simplifying idealizations of our
calculations. Nevertheless, the conclusion appears to be firm that plausible turbulent
magnetic fields are likely to have a significant effect on the magnetic field strengths that
are inferred from the observation of the circular polarization of spectral lines at radio
wavelengths. That is, their effect is unlikely to ``average out''. Clearly the exact ratio depends upon specific considerations outlined in the
foregoing. One-third is a representative estimate for the ratio
$(B_{\rm f} )_{\rm sd} /B_{\rm rms} $
of the rms deviations of the observationally inferred field to the
rms of the irregular (or turbulent) magnetic field that actually exists in the cloud. As
expected, there is a tendency for deviations from the average to cancel when the
contributions are summed along the paths of the rays of radiation. When the opacity
function $f(s)$ is a function of location $s$ because it depends upon matter density or other
considerations, $(B_{\rm f} )_{\rm sd} /B_{\rm rms} $
will naturally tend to increase. This increase can be viewed as a result of effectively reducing the number of elements along the path that
contribute which, in turn increases the scatter simply due to statistics. In MHD simulations, there should
be some correlation between the matter density and the strength of the magnetic field. This also tends
to increase $(B_{\rm f} )_{\rm sd} /B_{\rm rms} $, though we suspect that the effect
is small based on our examination of the correlation
between density and field strength. The results for the strong field (S) case in Figure 1
might seem to be in conflict with this conclusion. We suspect, however, that the relatively larger
values of $(B_{\rm f} )_{\rm sd} /B_{\rm rms} $
in this case are due mainly to the ``statistical'' effect of greater contrasts due to the
anisotropic variations in the density.

In a larger sense, the idealizations here are an incomplete description for star forming
clouds because gravity is absent in these idealizations. Gravitational attractions will tend to cause
clustering of matter, and hence more extreme variations in $f(s)$ than considered in our
calculations. In contrast, the results when $f(s)$ is constant in Figures 1 and 2 suggest that
purely statistical sampling provides velocity and magnetic fields that may be adequate to
understand the circular polarization of clouds where gravity is not a major effect and
when the finite resolution of the telescope is considered. This is not an altogether happy
conclusion since it potentially limits our prospects for identifying the physical
phenomena for which the non-thermal line breadths and variations in the magnetic fields
are indicators.

Our interpretation that the irregular component of the magnetic field
in the Orion A cloud is similar in magnitude to the average magnetic
field ($B_{\rm rms}= \left|\left\langle B_{\rm f}^{\rm obs}\right
\rangle\right|$) implies that the magnetic field is quite disordered
in this cloud. That the bulk motions of the gas are able to deform
the field to this degree tends to indicate that the turbulent kinetic
energy is at least comparable with the energy in the magnetic field,
and thus that the turbulent velocity is at least comparable with the
Alfven velocity. In numerical computations of MHD turbulence (including
the MHD fields that we are using), the
energy of the irregular component of the magnetic field tends to be
similar to the kinetic energy of the turbulence
(typically 30 to 60\%; V\'azquez-Semadeni et al. 2000). For Alfven waves,
these two components of the energy are equal. From this, it also follows
that the
turbulent and Alfven velocities should be comparable when $B_{\rm rms}=
\left|\left\langle B_{\rm f}^{\rm obs}\right\rangle\right|$. Padoan \&
Nordlund(1999) have emphasized that MHD computations in which the
turbulent velocities are super-Alfvenic lead to predicted features for
molecular clouds that are different in important ways from computations
with lower turbulent velocities. In view of the uncertainties in our
present analysis for $B_{\rm rms}$, we are unable to say whether the
turbulent velocities are somewhat larger or somewhat smaller than the
Alfven velocity. Alternatively, the ratio of the turbulent and Alfven
velocities could be obtained from the spectral linebreadth in Figure 3
and the average field strength in equation (14) if the mass density
were known reliably. However, as discussed in Section 3, the mass
density is poorly known. A mass density of about $10^4$ H-atoms
cm$^{-3}$ is required to yield an Alfven velocity that is equal to the
turbulent velocity. As discussed in Section 3, the Orion A cloud
probably is ``sheet-like''. At least along one line of sight, its molecular fraction is small. It may not, thus,  be representative of
star-forming, molecular clouds. We note that in other molecular clouds
where the matter density is better known, the velocities of the internal motions have
been interpreted as equal to the Alfven velocities (Crutcher 1999).

\acknowledgments

We are especially grateful to J. M. Stone for kindly making available
the results of the MHD simulations. We also acknowledge valuable contributions
by A. B. Men'shchikov in developing the initial computational
methods, and by H. W. Wyld in a number of helpful conversations. This work has
been supported by NSF Grants AST98-20641 and AST99-88104.

\clearpage

\end{document}